\documentclass[11pt]{article}
\usepackage{graphicx}
\usepackage{amssymb}
\usepackage{fancyhdr}
\fancypagestyle{plain}{ \fancyhf{} \cfoot{-~\thepage~-} }

\usepackage{amssymb}
\usepackage{mathrsfs}

\newcommand{\newsection}[1]{
\addtocounter{section}{1} \setcounter{equation}{0}
\setcounter{subsection}{0} \addcontentsline{toc}{section}{\protect
\numberline{\arabic{section}}{{\rm #1}}} \vglue .6cm \pagebreak[3]
\noindent{\bf  \thesection. #1}\nopagebreak[4]\par\vskip .3cm}
\newcommand{\newsubsection}[1]{
\addtocounter{subsection}{1}
\addcontentsline{toc}{subsection}{\protect
\numberline{\arabic{section}.\arabic{subsection}}{#1}} \vglue .4cm
\pagebreak[3] \noindent{\it \thesubsection.
#1}\nopagebreak[4]\par\vskip .3cm}

\renewcommand{\theequation}{\thesection.\arabic{equation}}

\newlength{\extraspace}
\newlength{\extraspaces}
\newcounter{dummy}
\newcommand{\bc}{\begin{center}}
\newcommand{\ec}{\end{center}}
\newcommand{\be}{\begin{equation}
\addtolength{\abovedisplayskip}{\extraspaces}
\addtolength{\belowdisplayskip}{\extraspaces}
\addtolength{\abovedisplayshortskip}{\extraspace}
\addtolength{\belowdisplayshortskip}{\extraspace}}
\newcommand{\ee}{\end{equation}}

\newcommand{\ba}{\begin{eqnarray}
\addtolength{\abovedisplayskip}{\extraspaces}
\addtolength{\belowdisplayskip}{\extraspaces}
\addtolength{\abovedisplayshortskip}{\extraspace}
\addtolength{\belowdisplayshortskip}{\extraspace}}
\newcommand{\ea}{\end{eqnarray}}

\newcommand{\ban}{\begin{eqnarray*}
\addtolength{\abovedisplayskip}{\extraspaces}
\addtolength{\belowdisplayskip}{\extraspaces}
\addtolength{\abovedisplayshortskip}{\extraspace}
\addtolength{\belowdisplayshortskip}{\extraspace}}
\newcommand{\ean}{\end{eqnarray*}}
\newcommand{\baa}{
\addtocounter{equation}{1} \setcounter{dummy}{\value{equation}}
\setcounter{equation}{0}
\renewcommand{\theequation}{\thesection.\arabic{dummy}\alph{equation}}
\begin{eqnarray}
\addtolength{\abovedisplayskip}{\extraspaces}
\addtolength{\belowdisplayskip}{\extraspaces}
\addtolength{\abovedisplayshortskip}{\extraspace}
\addtolength{\belowdisplayshortskip}{\extraspace}}
\newcommand{\eaa}{
\end{eqnarray}
\setcounter{equation}{\value{dummy}}
\renewcommand{\theequation}{\thesection.\arabic{equation}}}

\newcommand{\half}{\frac{1}{2}}
\newcommand{\del}{\partial}

\newcommand{\delb}{\bar{\del}}

\newcommand{\bt}{{\bf 10}}
\newcommand{\bfv}{{\bf 5}}
\newcommand{\bfb}{{\overline{\bf 5 \!}\,}}
\newcommand{\btb}{{\overline{\bf 10 \!}\,}}

\title{$F$-theory, GUTs and Chiral Matter}
\author{Martijn Wijnholt\\ \it \small Max Planck Institute, Potsdam-Golm \\ \\
June 18th, 2008
\\ \\
\it \small \sl Talk at the
Sixth Simons Workshop in Mathematics and Physics \\
\it \small \sl  Stony Brook University, June 16 - July 12, 2008 }
\date{}

\begin{document}

\maketitle


\newsection{Motivation}

This is a write-up of a two hour talk on
\cite{Donagi:2008ca},\cite{Donagi:2008kj}. The discussion is aimed
at non-experts and may be useful for people new to the subject.
Related work can be found in
\cite{Beasley:2008dc,Hayashi:2008ba,Beasley:2008kw}.\footnote{Many
interesting papers discussing phenomenological aspects of $F$-theory
have recently appeared; see
\cite{Heckman:2008qt,Marsano:2008py,Marsano:2008jq,Heckman:2008es,Buchbinder:2008at,Aparicio:2008wh,Collinucci:2008pf}.}

There are many ways to engineer Standard Model-like theories in
string theory. The class of models in this talk are based on two
principles, namely (1) local model building and (2) built-in
unification. Let us discuss these two principles in more detail.

(1) {\it Local model building}. Gauge groups are usually localized
on branes in string theory. Charged chiral matter arises from
intersections of these branes. This means that the matter we observe
at particle accelerators may all be localized in the extra
dimensions, and we may try to construct `local models' in which we
only study a small neighbourhood of the brane. What happens in the
rest of the extra dimensions is described by some unknown parameters
in the Lagrangian, whose values are set by the UV dynamics which we
have not yet included in our local description. We can state this
more precisely by saying that we would like the existence of a
decoupling limit $M_{pl,4}/\mu \to \infty$ while keeping
$g^2_{YM}(\mu)$ fixed.\footnote{This is actually too strong a
requirement and has to be relaxed a bit, since $U(1)$ couplings are
not asymptotically free and may have an interesting dependence on
the compactification. An interesting compactification effect on the
$U(1)$'s was discussed in \cite{Buican:2006sn}. Rather we will
assume such decoupling for all the couplings that are asymptotically
free.} This does not happen in generic models with branes. If we
scale up the volume of the internal space, we are typically also
forced to scale up the cycles on which the branes are wrapped,
turning off the gauge couplings.

Focussing on such local scenarios allows us to address questions of
particle physics without knowing the complicated dynamics of the
whole internal Calabi-Yau. It also alleviates the landscape problem
in that it should lead to much fewer and more predictive models.
However unfortunately it is not enough; in local models we can
engineer the MSSM with arbitrary parameters from open strings, as
well as any similar quiver model \cite{Wijnholt:2007vn}, and we will
likely never be able to rule out that these models cannot be
compactified. In order to get something more interesting, we need
some additional top-down input beyond requiring that the MSSM can be
realized in such a scenario. This is the purpose of the second
principle.

(2) {\it Built-in unification}. This means we would like to
construct GUT models with an $SU(5)$, $SO(10)$, or $E_6$ gauge
symmetry. One may even extend this list to $E_7$ and $E_8$ gauge
symmetry provided the symmetry is realized in higher dimensions and
broken in four dimensions. However such models can not be realized
using perturbative open strings:
\begin{itemize}
  \item gauge groups are
always $U(n)$,$SO(n)$ or $Sp(n)$, ruling out $E_6$;
  \item for $SO(10)$,
quarks and leptons sit in the ${\bf 16}$ of $SO(10)$ which is a
spinor representation, which again cannot be realized with
perturbative open strings;
  \item for $SU(5)$ the problem is that the top quark Yukawa
coupling uses the epsilon tensor of $SU(5)$, and so can only be
realized non-perturbatively, whereas the bottom quark Yukawa's are
realized perturbatively. However in Nature the top quark Yukawa is
of order one, and the bottom Yukawa is hierarchically smaller.
\end{itemize}

\begin{figure}[t]
\begin{center}
\renewcommand{\arraystretch}{1.5}
\begin{tabular}{|c|c|}
  \hline
  \quad {\it dim} \qquad & \qquad \qquad {\it stringy realization}\qquad \qquad \\
  \hline \hline
  10d \qquad & $E_8 \times E_8$ heterotic string \\
  9d & strongly coupled type I' \\
  8d & $F$-theory on ALE \\
  7d & $M$-theory on ALE \\
  6d & IIa on ALE/IIb with NS5 \\
  \hline
\end{tabular}\\[5mm]
\parbox{12cm}
{
{\bf Table 1}: \it Branes with exceptional gauge symmetry in string
theory.}
\renewcommand{\arraystretch}{1.0}
\end{center}
\end{figure}

On the other hand, in the old $E_8$ heterotic string there was no
problem with any of these issues. This is for purely group theoretic
reasons: the gauge indices of charged matter live in the coset
$E_8/G$, where $G$ is the GUT group, and the algebra of $E_8$ (and
also the other exceptional groups) allows for the spinor
representation of $SO(10)$ or the epsilon tensor in the up type
Yukawa coupling for $SU(5)$. So the upshot is we do not necessarily
need the heterotic string, but we do need branes with exceptional
gauge symmetry rather than conventional $D$-branes. These actually
appear in settings other than the heterotic string. In the table we
gave a list of branes with exceptional gauge symmetry in string
theory. However several requirements cut down this list further.
Five-branes are too much localized, they can always be separated in
the extra dimensions and thus have no charged chiral
matter.\footnote{This argument does not exclude getting chiral
matter from general 5d or 6d theories; eg. a 6d theory with both
vector and hypermultiplets would work.} The requirement of local
model building rules out the heterotic string and type I', and even
if we were interested in global models the type I' set-up doesn't
have any good constructive tools. A similar problem plagues the
$M$-theory models, and in addition the $M$-theory models have
phenomenological problems: they are still too much localized in that
even though they yield chiral matter, their Yukawa couplings have to
be generated through membrane instantons. Thus our two principles
lead us to consider the entry in the middle of the table:
$F$-theory. Fortunately, there are some powerful constructive
techniques available here from algebraic geometry, which we will
briefly review later.

So why were such $F$-theory models not pursued previously? After
all, brane models were constructed in IIb which is a close cousin,
and $F$-theory had also appeared in discussions on moduli
stabilization and Randall-Sundrum type scenarios. Essentially there
were two problems. The first is that no one had previously shown how
to engineer charged chiral matter in $F$-theory. There was also a
second serious problem, in that there seemed to be no good mechanism
for breaking the GUT group in $F$-theory. We will explain below how
these problems are addressed.


\newsection{Overview of model building with $F$-theory}

\newsubsection{Elliptic fibrations}

$F$-theory \cite{Vafa:1996xn} is basically a book-keeping device to
describe vacua of $IIb$ string theory with a varying axio-dilaton.
The $Sl(2,Z)$ duality group of IIb string theory acts as fractional
modular transformation on
\begin{equation}
\tau = a + i e^{-\phi}
\end{equation}
so that $\tau$ may be identified with the modular parameter of an
auxiliary torus (an `elliptic curve.') Thus we can formally attach
this torus at each point in the IIb space-time and speak of
twelve-dimensional compactifications of $F$-theory. Essentially this
is a clever change of variable: instead of specifying $\tau$
directly, which can get rather complicated due to the monodromies
acting on $\tau$, we specify the torus directly. The $T^2$ is
typically written in Weierstrass form, i.e. described as an equation
of the form
\be\label{Weierstrass} y^2 = x^3 + f x + g \ee
which can be done globally on the IIb space-time. Instead of
specifying $\tau$, we specify $f$ and $g$. The area of the torus has
no meaning in $F$-theory and should be taken zero. In these
variables, supergravity solutions with 7-branes (which have varying
dilaton) are much easier to describe, as we review next.

Let us label the one-cycles of the elliptic fiber as $a$ and $b$,
with $a \cap b = 1, a\cap a = b \cap b = 0$. On a subset of real
codimension 2 on the IIb space-time (namely $\Delta = 4f^3+27 g^2 =
0$) , the elliptic fiber pinches due to a 1-cycle $\gamma =p a + q
b$ shrinking to zero. As we go around this locus, which is called
the discriminant locus, the one-cycles undergo a monodromy following
the Picard-Lefschetz formula:
\begin{equation}
\delta \to \delta + (\delta \cap \gamma) \gamma
\end{equation}
Let us denote the holomorphic one-form on the $T^2$ by $\Omega$.
Then we can express $\tau$ as $\int_b \Omega/\int_a \Omega = \tau$.
With a little algebra, we see that the monodromy acts on $\tau$ as
\begin{equation}
\tau \to K_{[p,q]}\tau, \qquad K_{[p,q]}=\left(
\begin{array}{cc}
1+pq & p^2 \\
-q^2 & 1-pq \\
\end{array}
\right)
\end{equation}
We claim this identifies the locus with a $(p,q)$ 7-brane, i.e. a
type of 7-brane on which a $(p,q)$ string can end. To see this,
consider the case of a $(1,0)$ brane. In this case we have $\tau \to
\tau + 1$ as we go around a 7-brane, i.e. $a \to a+1$ and
$e^{-\phi}$ invariant. This is precisely the right monodromy for a
single $D7$-brane (it means that the 7-brane sources one unit of RR
flux). By applying $Sl(2,Z)$ duality transformations, we recover the
other cases.

Two $(p,q)$ 7-branes are said to be mutually local if
\begin{equation}
\left|
\begin{array}{cc}
p_1 & q_1 \\
p_2 & q_2 \\
\end{array}
\right| = 0
\end{equation}
Otherwise they are said the be mutually non-local. In the latter
case, the degrees of freedom on the branes are not independent of
each other and the combined system is generically strongly coupled.
In the former case, the degrees of freedom are independent and we
can always make the dilaton small and get a weakly coupled system.


There is another view on $F$-theory from $M$-theory, as follows.
$M$-theory on $T^2$, in the limit that the area goes to zero, is
equivalent to type IIb on a circle of radius $R = 1/A$, with
axio-dilaton given by the modular parameter of the $T^2$. By
fibering and taking $A \to 0$ we can get general $F$-theory
compactifications with 3+1 dimensional Poincar\'e invariance.

\newsubsection{Abelian gauge fields}

As usual in type II settings, abelian tensor fields on the brane
arise as zero modes of the ten-dimensional tensor fields localized
on the soliton. Since the gauge symmetry exists already in eight
dimensions, they must come form zero modes of $B_{NS}$ and $B_{RR}$.
However $B_{NS}$ and $B_{RR}$ are not invariant under the
monodromies; they form a doublet under the $Sl(2,Z)$ duality group.
In keeping with the philosophy of $F$-theory, we want to reformulate
this in terms of an object that can be specified globally, and is
not subject to monodromies. This can be done by encoding the
two-form fields in a three-form field:
\begin{equation}
C_{(3)} \sim (B_{RR} - \tau B_{NS})\wedge (dx - \tau dy) + c.c.
\end{equation}
where $x$ and $y$ are the two coordinates on the $T^2$. Note this
has two indices on the IIb space-time and one index on the elliptic
fiber. Three-form fields with different numbers of indices in the
base and the fiber do not exist in $F$-theory. This three-form field
$C_{(3)}$ is $Sl(2,Z)$ invariant and can be defined globally. By
compactifying on $S^1$ and going to $M$-theory, this corresponds to
the usual $C_{(3)}$ field of eleven-dimensional supergravity, except
that some components are turned off in the $F$-theory limit. The
four-form flux of this tensor field is conventionally called the
$G$-flux.

Now to get a 7-brane gauge field, we need to expand $C_{(3)}$ in
terms of real harmonic two-forms, with one index on the base and one
index on the fiber (so that the gauge field index lives in the IIb
space-time):
\begin{equation}
C_{(3)} = A_I \wedge \omega^I
\end{equation}
Consider for instance $F$-theory compactified on an elliptically
fibered $K3$-surface. There are 22 harmonic 2-forms, but one of
these has two indices on the base and one has two indices on the
fiber. Thus there are 20 harmonic forms we can expand in (which we
can further subdivide as $20 = 18 + 2$), which yields the same
number of $U(1)$ gauge fields as expected from the heterotic string
on $T^2$. Note that these harmonic forms cannot be normalizable in
the local supergravity solution for a 7-brane: if so we would get an
independent gauge field for each singular fibre, but there are 24
singular fibers in an elliptically fibered $K3$ and only 20 gauge
field from the 7-branes.

Later we will specialize to vacua with 3+1 dimensional Poincar\'e
invariance and $N=1$ supersymmetry. For this $F$-theory needs to be
compactified on an elliptically fibered Calabi-Yau four-fold and the
$G$-flux needs to satisfy certain conditions, which were derived by
Becker and Becker:
\begin{eqnarray}
 G &\in& H^{2,2}(CY_4) \qquad ({\rm F}-{\rm term\ equation}) \\ \nonumber
 J \wedge G &=& 0 \qquad \qquad \qquad ({\rm
 D}-{\rm term\
equation})
\end{eqnarray}
Here $J$ is the K\"ahler form on the Calabi-Yau. Note that these
conditions are similar to the ASD equations on the internal
worldvolume of a 7-brane, $F^{2,0}=0=J \wedge F^{1,1}$, and reduces
to them in a weak coupling limit.

We further want to specialize to local scenarios.  This can be
expressed by saying that the normal bundle to the cycle $S$ wrapped
by the gauge 7-branes should be negative, but let us instead use a
slightly weaker requirement. Deformations of the 7-branes correspond
to the parameters in $f$ and $g$ in (\ref{Weierstrass}), i.e they
are complex structure deformations of the Calabi-Yau four-fold.
These can be equivalently parametrized by harmonic $(3,1)$ forms,
and decomposing
\be \delta \Omega^{3,1} = \Phi_I \wedge \omega^I \ee
shows that they correspond to harmonic $(2,0)$ forms on $S$, the
four-cycle wrapped by the 7-brane. In order to eliminate such
massless adjoint fields, we need $h^{2,0} = 0$, which essentially
means that $S$ should be a Del Pezzo surface. (Hirzebruch surfaces
and the Enriques surface are also allowed by this argument, but less
phenomenologically interesting). A del Pezzo surface $dP_k$ is a
complex surface which can be constructed as a blow-up at $k\leq 8$
points on ${\bf P}^2$.

\newsubsection{Non-abelian gauge fields}

The non-abelian gauge bosons, as usual in type II settings, arise
from BPS states that can not be seen in supergravity and have to be
added by claiming some knowledge of the UV completion.

 \begin{figure}[t]
\begin{center}
            \scalebox{.6}{
               \includegraphics[width=\textwidth]{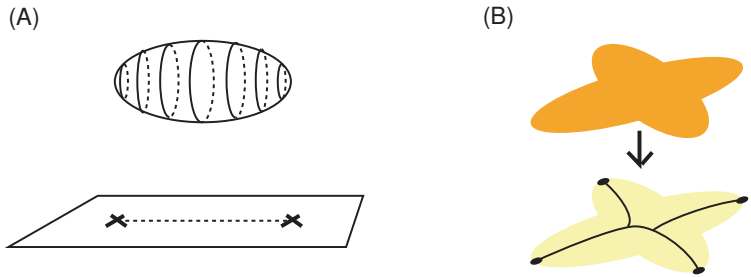}
               }
\end{center}
\vspace{-.5cm} \begin{center} \parbox{12cm}{\caption{ \it An open
fundamental string in type IIb lifts to a membrane wrapping an
exceptional cycle in $F$-theory. Some W-bosons may correspond to
ground states of multi-pronged $(p,q)$ strings .}}\end{center}
\label{VanishingW}
\end{figure} 

Let us consider two parallel D7-branes. The non-abelian gauge bosons
that enhance the symmetry to $SU(2)$ come from open strings
stretched between the two 7-branes. How is this seen in $F$-theory?
Consider the path associated to an open string stretching between
the branes. On top of each point of this path we can associate a
1-cycle of the $T^2$ fiber, which we take to be the $(1,0)$ cycle.
On the left and right ends of the path, this $(1,0)$ cycle shrinks
to zero. Altogether then we reconstruct a topological $S^2$. Using
the $M$-theory perspective, we can wrap an $M2$-brane on this $S^2$,
which turns into the fundamental open string as we go to $F$-theory.
As the we let the 7-branes approach each other, the $S^2$ shrinks to
zero and the Calabi-Yau fourfold develops an $A_1$ singularity. (The
IIb space-time is still perfectly smooth, only when we add the
elliptic fibration do we see the singularity). As the $S^2$ shrinks
to zero, the ground states of the wrapped $M2$ brane or open
fundamental string become massless, yielding the off-diagonal
components of an $SU(2)$ vector multiplet. We get both $W^+$ and
$W^-$ by reversing the orientation of the membrane.

\begin{figure}[t]
\begin{center}
\renewcommand{\arraystretch}{1.5}
\begin{tabular}{|c|c|c|c|c|}
\hline

 {\it ord(f)} & {\it ord (g)} & {\it ord($\Delta$)}   &
                               {\it fiber type} & {\it singularity type} \\
                               \hline \hline
$\geq 0$ & $\geq 0$ & $0$ & smooth & $-$\\
$0$ & $0$ & $n$ &  $I_n$  & $A_{n-1}$ \\
$\geq 1$ & $1$ & $2$& $II$ & $-$\\
$1$ & $\geq 2$ & $3$ &  $III$ & $A_1$ \\
$\geq 2$ & $2$ & $4$ &   $IV$  & $A_2$\\
$2$ & $\geq 3$ & $n+6$ &  $I_n ^*$ & $D_{n+4}$ \\
$\geq 2$ & $3$ & $n+6$ &  $I_n ^*$ & $D_{n+4}$ \\
$\geq 3$ & $4$ & $8$ &  $IV^*$ & $E_6$ \\
$3$ & $\geq 5$ & $9$ &  $III^*$  & $E_7$\\
$\geq 4$ & $5$ & $10$ &  $II^*$  & $E_8$ \\ \hline
\end{tabular}\\[5mm]
\parbox{10cm}
{ \bf Table 2: 
\it Kodaira classification of singularities of elliptic fibrations,
indicating the order of vanishing of $\Delta$, $f$ and $g$.}
\renewcommand{\arraystretch}{1.0}
\end{center}
\end{figure}

\begin{figure}[t]
\begin{center}
            \scalebox{.5}{
               \includegraphics[width=\textwidth]{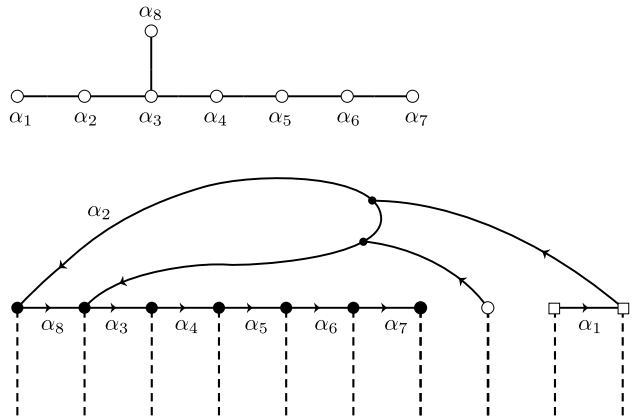}
               }
\end{center}
\vspace{-.5cm} \caption{ \it Representation of the fundamental roots
of $E_8$ from \cite{DeWolfe:1998zf}. Here a cross denotes an
$A$-brane, a circle denotes a $B$-brane, and a box denotes a
$C$-brane.}\label{E8roots}
\end{figure}

The type of singularities in an elliptic fibration were classified
by Kodaira (see table 2). The elliptic fibration may develop an ADE
singularity by letting various branes approach each other, and one
would naturally expect that by wrapping $M2$ branes on the vanishing
cycles one gets an enhanced ADE gauge symmetry. How do we understand
these more general situations from the IIb space-time? It turns out
that the exceptional cycles of an ALE do not necessarily project to
open strings with two ends, but may yield so-called multi-pronged
strings with multiple ends. This is the key to getting the
exceptional groups and can happen when the dilaton cannot be taken
small. All the ADE Lie algebras have been reproduced from such
(generally multi-pronged) strings. For instance the roots of $E_8$
can be recovered from a configuration of seven $A$-branes, one $B$
brane, and two $C$-branes, where $A=(1,0), B=(1,-1), C=(1,1)$, see
figure \ref{E8roots}. Configurations for the type D Lie algebras are
similar except that they just use one $C$ brane instead of two. A
$B$-brane and $C$-brane can be combined into an orientifold plane
and yield weak coupling limits, but this is not possible for the
exceptional cases.

\newsubsection{Charged matter}

Now we'd like to understand how to get charged matter. This should
arise form the intersections of 7-branes. When the discriminant
locus self-intersects, the order of vanishing of $\Delta$ increases
and the singularity type of the fibration will be enhanced, i.e.
there are extra vanishing cycles sitting over the intersection of
the 7-branes. By wrapping membranes on them and quantising, we get
six-dimensional hypermultiplets living on the intersection,
basically because that's the only possibility given the symmetries.
The hypermultiplets naturally sit in a kind of generalized
bifundamental representation of the gauge groups on the 7-branes.

 \begin{figure}[t]
\begin{center}
            \scalebox{.4}{
               \includegraphics[width=\textwidth]{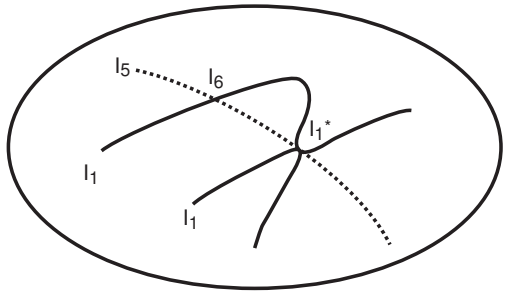}
               }
\end{center}
\vspace{-.4cm} \caption{ \it An $I_5$ locus in the IIb space-time,
intersecting an $I_1$ locus over an $I_6$ degeneration and an
$I_1^*$ degeneration.}\label{IntersectingBranes}
\end{figure} 

Let us consider some examples that are relevant for GUT model
building. We take an $I_1$ locus (which has a single pinched cycle
on the $T^2$) with an $I_5$ locus (which has an $A_4$ singularity,
hence an $SU(5)$ gauge group; the $T^2$ has degenerated to five
$S^2$'s intersecting according to the affine $A_4$ Dynkin diagram).
Over the intersection the singularity can get enhanced in two ways:
either to $I_6$, which has a $SU(6)$ Lie algebra worth of vanishing
cycles, or an $I_1^*$ singularity, which has a $SO(10)$ Lie algebra
worth of vanishing cycles. Decomposing
\begin{equation}
Ad(SU(6)) = Ad( SU(5)) + \bfv + \bfb + {\bf 1}
\end{equation}
we see that the membranes wrapped on  the extra vanishing cycles
naturally yield a hypermultiplet in the $\bfv$, as is familiar from
perturbative D-branes.  Similarly decomposing
\begin{equation}
Ad(SO(10)) = Ad( SU(5)) + \bt + \btb + {\bf 1}
\end{equation}
we see that we naturally get a hypermultiplet in the $\bt$ on the
intersection. As another example, consider the intersection of an
$I_1$ and an $I_1^*$ singularity, which yields an enhancement to
$E_6$. Decomposing
\begin{equation}
Ad(E_6) = Ad( SO(10)) + {\bf 16} + \overline{\bf 16} + {\bf 1}
\end{equation}
we see that the `bifundamental' is in this case a hypermultiplet in
the spinor representation, which we could not get in perturbative
IIb. It should be noted that intersections of branes which are not
mutually local are highly non-transversal.

All this was known in the nineties. On the other hand, before
\cite{Donagi:2008ca} it was not understood how to get chiral matter
in $F$-theory. Of course this is a crucial issue for model building
because quarks and leptons are chiral.

To understand how to get chiral matter, one may ask how we get
chiral matter in type IIb string theory. This was also already
understood in the nineties, but somehow did not make its way into
the $F$-theory literature. The intersection of the 7-branes is six
dimensional, so we have two dimensions left to make the
hypermultiplet chiral. We compactify on a Riemann surface and write
the six-dimensional hypermultiplet spinors as $\psi \otimes \chi$
where $\chi$ are four-dimensional spinors and $\psi$ are spinors on
the Riemann surface. The Dirac action splits in a four-dimensional
piece and a piece for the fermions on the Riemann surface:
\begin{equation}
\int_\Sigma \psi^{\bar{a}}_+ \delb_{A_1 - A_2} \psi^{{a}}_+ +
\psi^{\bar{a}}_- \del_{A_1-A_2} \psi^{{a}}_-
\end{equation}
Here $a$ corresponds to the $(N_1, \bar{N}_2)$ and $\bar{a}$
corresponds to the $(\bar{N}_1,N_2)$. When no flux is turned on, we
have two fermionic zero modes $\psi^a_+, \psi^{\bar{a}}_+$ and their
complex conjugates. The spinors of a six-dimensional hypermultiplet
may be decomposed as
\begin{equation}
\Psi^a = \chi_\alpha \otimes \psi^a_+ + \bar \chi_{\dot{\alpha}}
\otimes \psi^a_-, \qquad \Psi^{\bar{a}} = \bar{\chi}_{\dot{\alpha}}
\otimes \psi^{\bar{a}}_- +   \chi_\alpha \otimes \psi^{\bar{a}}_+
\end{equation}
Here $\chi_\alpha$ is a four-dimensional Weyl spinor, $\psi$ are the
complex spinors on $\Sigma$. Thus to get chiral matter we want to
have zero modes for $\psi^a_+,\psi^{\bar{a}}_-$, but not for
$\psi^a_- , \psi^{\bar{a}}_+$. The way to do this is familiar from
the Schwinger model of two-dimensional electro-dynamics: since the
fermions do not transform in a real representation of the gauge
group, we can introduce an asymmetry by turning on a background flux
$F=F_1 - F_2$ through $\Sigma$, and the net number of chiral fermion
zero modes will then be given by
\begin{equation}
N_{gen} = {1\over 2\pi} \int_\Sigma F
\end{equation}
(One can actually do better in situations with $N=1$ SUSY and
compute the absolute number, rather than just the net number, but we
will not discuss that here). Clearly this must also be true in
$F$-theory. Thus our main task consists of two things: (1) we need
to formulate this in $F$-theory language, by using the $G$-flux; and
(2) we need to generalize it to branes which are not mutually local.
And finally of course we need to implement it in explicit models.

As for the first, let us consider two stacks of intersecting
$D$-branes. The overall $U(1)$'s of these two stacks are encoded as
\be G \sim F_1 \wedge \omega_1 + F_2 \wedge \omega_2 \ee
Now let's consider a fundamental string stretching between these two
stacks, and lift it to a membrane wrapped on a vanishing cycle
$\alpha$. This is one of the extra vanishing cycles which sits over
the intersection, and corresponds to a root of $SU(N_1 + N_2)$ which
is not in $SU(N_1) \times SU(N_2)$. Then
\be  \int_\alpha \omega_1 = +1, \qquad \int_\alpha \omega_2 = -1 \ee
since the integral of $\omega$ over $\alpha$ gives the charge of the
wrapped membrane under the $U(1)$ associated to $\omega$. Now we can
construct a four-cycle $\Sigma \times \alpha$, and we can write the
net number as
\be N\ =\ {1\over 2 \pi}\int_\Sigma F_1 -F_2\ =\ {1\over 2\pi}
\int_{\Sigma \times \alpha} G \ee
The last expression can be used generally, even when there exist no
harmonic forms in the class of $\omega_1$, $\omega_2$ or even
$\omega_1-\omega_2$. (It is in fact crucial in GUT models that such
$\omega$ do not exist because we don't want extra massless $U(1)$'s
in the low energy spectrum, but we do want their fluxes in order to
get chiral matter).

For the second, we can use the same procedure of integrating over
suitable vanishing cycles, even though the charges of the extra
states may not be $\pm 1$. Eg. in the case of an $I_5$ and $I_1$
locus intersecting over $I_1^*$, we get
\be N\ =\ {1\over 2\pi} \int_\Sigma 2F\ =\ {1\over 2\pi}
\int_{\Sigma \times \alpha} G \ee
because the extra roots (similar to the root $\alpha_2$ in figure
\ref{E8roots}) carry charge $\pm 2$ under the overall $U(1)$'s
associated to the intersecting 7-branes.

It is interesting to compare this with the heterotic string. Here we
have a Calabi-Yau three-fold $Z$ and a holomorphic bundle $V$ which
breaks the $E_8$ gauge group in ten dimensions. There is a famous
formula for the number of generations in that context:
\begin{equation}\label{}
    N_{gen} = \half \int_Z c_3(V)
\end{equation}
As shown in \cite{Donagi:2008ca}, under $F$-theory/heterotic duality
this turns precisely into the $F$-theory expression for the net
number.

\newsubsection{Explicit models}

To conclude, we still need a method for constructing models.
Constructing local Calabi-Yau four-folds was explained by
Freedman-Morgan-Witten, and turns out to be remarkably easy. Our
four-fold will be an ALE fibration over a del Pezzo surface $S$,
which we get to choose. We write the equation of a deformed $E_8$
singularity as
\be\label{su5model}  y^2 = x^3 + a_0 z^5 + a_2 x z^3 + a_3 y z^2 +
a_4 x^2 z + a_5 xy \ee
This can easily be embedded in an elliptic fibration
(\ref{Weierstrass}) by adding some extra terms which are subleading
at the singularity. Here the $a_i$ are certain polynomials on the
del Pezzo surface $S$, i.e they are sections of line bundles on $S$.
Note that when all $a_i=0$ except for $a_0$, we have an $E_8$
singularity, but when $a_5 \not = 0$ we generically have an $A_4$
singularity, so this corresponds to an $E_8$ gauge theory which is
broken to an $SU(5)$ GUT model through compactification. To specify
the ALE fibration, we need to specify a class $\eta \in H^2(S)$, and
we need to choose five polynomials with Chern classes
\be
 a_i \sim \eta- i\,c_1(TS)
\ee
This completely specifies the local geometry, i.e. the 7-branes and
their intersections. Of course $\eta$ should be sufficiently
positive so that the sections $a_i$ exist. In addition we need to
specify a flux in order to get chiral matter. These fluxes are
specified by a (half)-integer $\lambda$. Due to a formula of
Curio's, the net amount of chiral matter is given by
\be N_{gen} = \lambda \,\eta \cdot_{dP} (\eta - 5 c_1) = \lambda
\int_{\Sigma_{\bt}} \eta \ee
where $\Sigma_\bt = \{ a_5 = 0\}$ is the matter curve on which the
hypermultiplet in the $\bt$ is localized. For the case of $SU(5)$
models, $\lambda$ should be of the form $\half + $integer. It's an
easy exercise to make your own three-generation model using these
formulae. Some examples may be found in
\cite{Donagi:2008ca}.\footnote{In v1 of \cite{Donagi:2008ca} it is
incorrectly stated that this flux is not primitive. An error in the
expression for the flux was pointed out to us by Taizan Watari.}

\newsection{Breaking the GUT group}

We succeeded in constructing four-dimensional GUT gauge groups and
chiral matter. However this is not quite what we want: at low
energies we only observe the Standard Model gauge group,
$SU(3)\times SU(2)\times U(1)$. We still need a mechanism for
breaking the GUT group, but in such a way that we don't screw up the
nice predictions like gauge coupling unification. There are
basically three ideas for doing this:

\begin{enumerate}
  \item {\it Adjoint Higgses}. This is a priori allowed in
  $F$-theory but yields a conventional four-dimensional GUT model
  with all the associated problems. In addition, this would not be a
  local model because the existence of an adjoint means that the
  7-brane is allowed to explore the whole internal space.
  \item {\it Discrete Wilson lines}. This is the primary method for
  breaking the GUT group in the heterotic string, or on manifolds of
  $G_2$ holonomy in $M$-theory. However in local models in
  $F$-theory, the seven-brane is necessarily wrapped on a four-cycle
  with trivial fundamental group, so this method is not available
  for us.
  \item {\it $U(1)$ fluxes with a hypercharge component}. This
  method is a priori also available in the heterotic string, but
  turns out to spoil unification. However, due to a mechanism
  discovered in \cite{Buican:2006sn}, it turns out this option is available in
  $F$-theory.
\end{enumerate}

\begin{figure}[t]
\begin{center}
            \scalebox{.3}{
               \includegraphics[width=\textwidth]{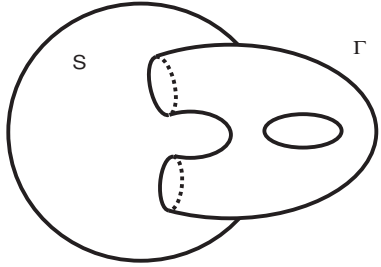}
               }
\end{center}
\vspace{-.5cm} \caption{ \it If we thread fluxes through cycles
which are trivializable in the bulk, then the overlap with zero
modes of bulk axions vanishes.}\label{Trivialization}
\end{figure}

Let us elaborate on the last mechanism. If we turn on an internal
flux for hypercharge then the gauge group will break to the
commutant of $Y$ in $SU(5)$. This is precisely the Standard Model
gauge group $SU(3) \times SU(2) \times U(1)$, and to leading order
the $SU(5)$ relations between the gauge couplings will be preserved,
so at first sight this looks promising. So why is this not usually
claimed to be the solution to breaking the GUT group in the
heterotic string? This is because in the heterotic string there is a
coupling
\begin{equation}\label{}
    \int d^{10}x\, (dB + A \wedge F)^2
\end{equation}
As a result of this coupling, if we turn on an internal hypercharge
flux, the four-dimensional hypercharge field will swallow an axion
(a zero mode of the $B$-field) and pick up a mass. Analogously in
$F$-theory there is a Chern-Simons coupling to RR forms:
\begin{equation}\label{}
    \mathscr{L} \sim \int C_4 \wedge G \wedge G
\end{equation}
Now we expand
\begin{equation}\label{}
    G = F_Y \wedge \omega^Y + G_{int}, \qquad C_4 = C_2^M \wedge \beta_M
\end{equation}
where $\beta^M$ is a basis for $H^2(B_3)$, $B_3$ is the base of the
Calabi-Yau fourfold, and $G_{int}$ is the internal $G$-flux which
has all its indices on the four-fold. This leads to
\begin{equation}\label{}
   \mathscr{L} \sim  \Pi^Y_M\int d^4 x\,  F_Y \wedge C_2^M, \qquad \Pi_M^Y = \int_{CY_4}
    \beta_M \wedge \omega^Y \wedge G_{int}
\end{equation}
This is a St\"uckelberg coupling for the hypercharge gauge field. So
we are in danger of generating a mass for hypercharge, which is
certainly not what we want. If we turn on an internal hypercharge
flux, then we have
\begin{equation}\label{}
    \Pi_M^Y \sim \int_S i^*\beta_M \wedge F_Y
\end{equation}
Now in $F$-theory we have a possibility that we didn't have in the
heterotic string. In $F$-theory, $\Pi_M^Y=0$ for all $M$ does not
imply that the internal hypercharge flux must be zero. Instead it is
equivalent to the statement that the Poincar\'e dual two-cycle $\Xi$
of $F_Y$ in $S$ becomes a boundary when embedded in $B_3$, see
figure \ref{Trivialization}. I.e. it is a topological condition on
the compactification of the local model. This happens quite
generically, but not in $F$-theory duals of the heterotic string.
Essentially the same mechanism for getting a massless hypercharge in
models with branes at singularities was first discovered in
\cite{Buican:2006sn}.

This mechanism for breaking the GUT group also does not lead to a
doublet-triplet splitting problem. If we arrange the fluxes
correctly, there are basically no four-dimensional $SU(5)$ partners
of the Higgs doublets. In this aspect it is similar to breaking by
discrete Wilson lines.

Now a pure hypercharge flux actually turns out to give the wrong
spectrum. However there are additional (massive) $U(1)$ symmetries
available from the flavour branes, which commute with $SU(5)_{GUT}$.
So instead one has to turn on a flux for a massive $U(1)$ symmetry,
which is not hypercharge but has a hypercharge component. The fact
that we may turn on such a flux may sound strange, but one may check
\cite{Donagi:2008kj} that the equations of motion (which require the
$G$-flux to be harmonic, primitive and of type $(2,2)$) can in fact
be satisfied. We refer to \cite{Donagi:2008kj} for further details.

\newsection{Phenomenological signatures}

I'd like to briefly discuss three signatures of the models we have
described. They are: monopoles, threshold corrections to
unification, and proton decay.

\newsubsection{GUT monopoles}

One of the classic predictions of conventional unification models is
the existence of monopoles carrying hypercharge. Let us recall the
argument. We have the long exact homotopy sequence
\begin{equation}
\ldots \to \pi_2(G) \to \pi_2(G/H) \to \pi_1(H) \to \pi_1(G) \to
\ldots
\end{equation}
Here $G$ is the GUT group and $H$ is what's left after breaking. If
we use an adjoint Higgs to break $SU(5)$, then we have
\begin{equation} H = [SU(3) \times SU(2) \times U(1)]/Z_6
\end{equation}
Monopoles are classified by $\pi_2(G/H)$. Since $\pi_2(G) =\pi_1(G)=
0$, we see that they can be equivalently classified by $\pi_1(H)$
(which corresponds to the monodromy around a Dirac string). Thus
these monopoles carry hypercharge, as well as an additional $Z_3$
colour and $Z_2$ electro-weak charge.

How do we find such monopoles in our GUT models? Recall that the
Poincar\'e dual of the GUT breaking flux on the del Pezzo surface is
a class $\Xi$ which is the boundary of a 3-chain $\Gamma$ in the IIb
space-time. One can wrap a $D3$-brane on $\Gamma$ and check that
when we turn on an internal flux for hypercharge, the resulting
four-dimensional particle carries magnetic hypercharge.

In addition there are other solitons: strings from $D3$ branes
wrapped on 2-cycles of the del Pezzo, domain walls interpolating
between vacua with and without $G$-flux. It might be interesting to
explore their phenomenological consequences

\newsubsection{Threshold corrections}

I've argued that there is a mechanism for breaking the GUT group
that preserves the $SU(5)$ relations between the gauge couplings.
However there are many heavy states charged under $SU(5)$ and
integrating them out will give small corrections to the couplings at
the GUT scale. There are basically two types of corrections:
\begin{enumerate}
  \item loop corrections from KK modes of the eight-dimensional
  gauge theory. These can be expressed as Ray-Singer torsion of the
  compactification manifold and matter curves;
  \item contributions from the massive excitations of open strings that were not included
  in the eight-dimensional gauge theory (analogous to `gravitational smearing' in 4D GUT models).
  Fortunately the leading
  contributions are constrained by non-renormalization conjectures.
\end{enumerate}

The leading corrections can be computed/estimated. They are of order
a few percent and come with varying signs, and so can be consistent
with the known values of the couplings at low energies. See
\cite{Donagi:2008kj} for details.

\newsubsection{Proton decay}

Generic GUT models have catastrophic proton decay.  The basic
problem is that the leptons and Higgses have the same quantum
numbers, so that if we have down type couplings
\begin{equation}
\bt_m \cdot \bfb_m\cdot \bfb_h \end{equation} %
then we also expect $R$-parity violating couplings
\begin{equation}
\bt_m \cdot \bfb_m\cdot \bfb_m \end{equation} %
which lead to proton decay. Even if these are absent by $R$-parity,
exchange of higgsino colour triplets leads to baryon number
violating dimension five operators $d^2\theta QQQL$ and $d^2\theta
UDUE$. The classic signature for this type of decay is $p \to K^+
\bar\nu$. With our mechanism for breaking the GUT group, there are
no four-dimensional colour $SU(5)$ partners of the Higgsinos,
however there is a tower of Kaluza-Klein modes with the same quantum
numbers.

The basic idea for eliminating dimensions four and five operators is
to use localization of the wave functions in the extra dimensions.
Such ideas appeared in the phenomenology literature around '99 and
in the context of $F$-theory unification models in
\cite{Tatar:2006dc}. (In this paper the precise $F$-theory
description of chiral matter and couplings was not yet understood,
but the qualitative picture was deduced using $F$-theory/heterotic
duality.) The ideas of \cite{Tatar:2006dc} can be generalized
slightly (and a general prescription for $E_8 \to SU(5)_{GUT}$
models in (\ref{su5model}) is given in \cite{Donagi:2008kj}) but for
the purpose of this talk, let me assume that the dimension four and
five operators are eliminated and move on to the dimension six
operators.

The dimension six proton decay operators are mediated by massive
gauge bosons in the representation $(2,3)_{-5/6}$, leading to the
decay $p \to \pi^0 e^+$. These are KK modes of the eight-dimensional
gauge fields, i.e. they are bulk modes and so we cannot appeal to
localization of the wave functions to reduce their effect. Of course
such operators are suppressed by $1/M_{GUT}^2$ so we would not
necessarily expect them to lead to any problems. The amplitude is
hard to calculate exactly because it depends on the profile of the
zero modes as well as the Green's function for the Laplacian on $S$.
However one may get the parametric dependence. The leading term in
the limit $\alpha_{GUT} \to 0$ turns out to be\footnote{This differs
from the estimate given in v1 of \cite{Donagi:2008kj}, which was a
bit too naive.}
\be \mathscr{M} \sim \alpha_{GUT} \log \alpha_{GUT}^{-1} {J_\mu
\tilde{J}^\mu(0)\over M_{GUT}^2 }\ee
where $J^\mu = \bar{\psi} \gamma^\mu \psi$ and $\psi$ corresponds to
the $\bt$ or $\bfb$. The analogous amplitude in four-dimensional GUT
models is
\be \mathscr{M} \sim \alpha_{GUT}  {J_\mu \tilde{J}^\mu(0)\over
M_{GUT}^2} \ee
so there is a mild parametric enhancement of proton decay in
$F$-theory. In practice with $\alpha_{GUT} \sim 1/25$ this is not so
big however, and it is not clear how the numerical coefficient
compares with that of four-dimensional models, so the best guess is
that there is only a minor enhancement and such decay won't be seen
any time soon.

\newsection{Outlook}

The $F$-theory models we described here seem to be the first new
class of models since the heterotic string that can successfully
incorporate unification. Two intriguing aspects are the option of
constructing local models (which allows for scenarios not available
in the heterotic string, like gauge mediation), and the method for
breaking the GUT group using fluxes. Despite some crucial
differences, there are also many similarities with the heterotic
models and also with models based on $G_2$ manifolds (assuming they
exist, which has not yet been shown). We expect that exploring the
relations between these models will lead to many new insights.
Further, if supersymmetric unification still holds up after the LHC
then with the state of knowledge today the $F$-theory models look
like some of the most viable phenomenological candidates.

\renewcommand{\Large}{\normalsize}

 \end{document}